\documentclass[english,aps,manuscript]{revtex4}
\usepackage[T1]{fontenc}
\usepackage[latin9]{inputenc}
\usepackage{textcomp}

\makeatletter
\@ifundefined{textcolor}{}
{%
 \definecolor{BLACK}{gray}{0}
 \definecolor{WHITE}{gray}{1}
 \definecolor{RED}{rgb}{1,0,0}
 \definecolor{GREEN}{rgb}{0,1,0}
 \definecolor{BLUE}{rgb}{0,0,1}
 \definecolor{CYAN}{cmyk}{1,0,0,0}
 \definecolor{MAGENTA}{cmyk}{0,1,0,0}
 \definecolor{YELLOW}{cmyk}{0,0,1,0}
}

\makeatother

\usepackage{babel}
\begin{document}

\title{Structure formation in the presence of relativistic heat conduction:
corrections to the Jeans wave number with a stable first order in
the gradients formalism }

\author{J. H. Mondragón-Suárez$^{1}$, A. Sandoval-Villalbazo$^{1}$, A.
L. García-Perciante,$^{2}$}

\address{$^{1}$ Depto. de Física y Matemáticas, Universidad Iberoamericana,
Prolongación Paseo de la Reforma 880, México D. F. 01219, México.
\linebreak{}
$^{2}$Depto. de Matemáticas Aplicadas y Sistemas, Universidad Autónoma
Metropolitana-Cuajimalpa, Artificios 40 México D.F. 01120, México.}
\begin{abstract}
{\normalsize The problem of structure formation in relativistic dissipative
fluids was analyzed in a previous work within Eckart\textquoteright{}s
framework, in which the heat flux is coupled to the hydrodynamic acceleration,
additional to the usual temperature gradient term. It was shown that
in such case, the pathological behavior of fluctuations leads to the
disapperance of the gravitational instability responsible for structure
formation \cite{GRGhumo}. In the present work the problem is revisited
now using a constitutive equation derived from relativistic kinetic
theory. The new relation, in which the heat flux is not coupled to
the hydrodynamic acceleration, leads to a consistent first order in
the gradients formalism. In this case the gravitational instability
remains, and only relativistic corrections to the Jeans wave number
are obtained. In the calculation here shown the non-relativistc limit
is recovered, opposite to what happens in Eckart's case \cite{HL}.}{\normalsize \par}
\end{abstract}
\maketitle

\section{{\normalsize Introduction }}

Relativistic hydrodynamics has been a subject of interest since the
1940\textasciiacute{}s. However, many open questions remain unanswered,
in particular the nature of heat conduction in hot gases has recently
been under research and debate. The main issue in such a subject concerns
the particular structure of the constitutive equation for dissipative
energy flow in the special relativistic case. On the one hand, C.
Eckart in 1940, proposed a relation which coupled heat to the hydrodynamic
acceleration \cite{Eckart}. This relation was obtained phenomenologically
as a sufficient condition for the entropy production to remain positive.
Later in 1985, Hiscock and Lindblom working relativistic dissipative
fluids, found generic instabilities in Eckart\textquoteright{}s formalism
which in part lead to the developement and use of extended theories
\cite{HL}. A subsequent study showed that these instabilities are
due to the presence of the hydrodynamic acceleration in the heat flux
constitutive equation \cite{GRG09}. 

On the other hand several authors have found that kinetic theory predicts
a coupling of heat with gradients of state variables instead of the
acceleration \cite{degrootR,ck,PhyA}. At the present time, both type
of constitutive equations have been used in order to study the Rayleigh-Brillouin
scattering spectrum, finding that by introducing Eckart\textquoteright{}s
equation the results were inconsistent with the observations, while
the use of the equations in terms of gradients eliminates the inconsistencies
\cite{pre09}.

In recent work, two of us addressed the question of structure formation
in relativistic dilute gases in the presence of dissipation introducing
Eckart\textquoteright{}s equation for the heat flux \cite{GRGhumo}.
The results obtained were similar to those presented by Hiscock \&
Lindblom, that is the apperance of an instability of early onset triggered
by the fluid itself. This instability, additional to not being observed,
rules out any possibility of structure formation in such gases since
the growth of density fluctuations due to the gravitational field
are overriden. 

In the present work, we instead use the consitutive equation obtained
in relativistic kinetic theory in the study of structure formation
and find that generic instabilities are eliminated in such a way so
that gravity becomes dominant. We finally establish the relativistic
corrections to the critical wave number. 

We have structured this paper as follows. In Sect. II we establish
the system of relativistic transport equations for a gas including
gravitational effects within the Newtonian limit and linearize around
equilibrium values for the state variables in order to obtain a set
of linear equations for the fluctuations. In Sect. III we calculate
the dispersion relation for the system in the Fourier-Laplace space
and infer from its solutions the behavior of density fluctuations.
Section IV is devoted to the explicit calculation of the relativistic
correction to the Jeans wave number in a dissipative medium and compare
the result with its non-relativistic counterpart. Conclusions and
final remarks are included in Sect. V.

\section{{\normalsize Linearized Transport Equations }}

The behavior of density fluctuations in a fluid is analyzed by studying
the response of the corresponding linearized system of transport equations
when the state variables are perturbed from their equilibrum values
\cite{bernepecora}. That is, if $X$ is a state variable one considers
\begin{equation}
X=X_{0}+\delta X\label{eq:1}
\end{equation}
where $X_{0}$ is the equilibrium value and $\delta X$ the fluctuation.
The gas here considered is a dilute cloud of particles with a temperature
high enough, so that the relativistic effects in the molecular motion
is relevant. We consider the temperature $T$, number density $n$
and hydrodynamic four-velocity $u^{\mu}$ as the independent, local
equilibrium, state variables. The system of equations where the hypothesis
in Eq. (\ref{eq:1}) is to be introduced is composed of two tensor
conservation equations namely,
\begin{equation}
N_{;\mu}^{\mu}=0\label{eq:2}
\end{equation}
and
\begin{equation}
T_{;\nu}^{\mu\nu}=0\label{eq:3}
\end{equation}
where
\begin{equation}
N^{\mu}=nu^{\mu}\label{eq:4}
\end{equation}
is the particle four-flow and
\begin{equation}
T_{\nu}^{\mu}=\frac{n\varepsilon}{c^{2}}u^{\mu}u_{\nu}+ph_{\nu}^{\mu}+\pi_{\nu}^{\mu}+\frac{1}{c^{2}}q^{\mu}u_{\nu}+\frac{1}{c^{2}}u^{\mu}q_{\nu}\label{eq:5}
\end{equation}
is the stress-energy tensor. Here $\varepsilon$ is the energy density
per particle, $c$ the speed of light and $p$ the hydrostatic pressure.
The dissipative fluxes are the Navier tensor $\pi^{\mu\nu}$ and the
heat flux $q^{\mu}$. We have also used the standard expression for
the spatial projector
\[
h^{\mu\nu}=g^{\mu\nu}+\frac{u^{\mu}u^{\nu}}{c^{2}}
\]
in a $+++-$ metric and the orthogonality relations
\[
u^{\nu}q_{\nu}=0\qquad u^{\nu}\pi_{\mu\nu}=0
\]
are also assumed. Thus, the transport equations for the relativistic
fluid may be written as
\begin{equation}
\dot{n}+n\theta=0\label{eq:6}
\end{equation}
\begin{eqnarray}
\left(\frac{n\varepsilon}{c^{2}}+\frac{p}{c^{2}}\right)\dot{u}_{\nu}+\left(\frac{n\dot{\varepsilon}}{c^{2}}+\frac{p}{c^{2}}\theta\right)u_{\nu}+p_{,\mu}h_{\nu}^{\mu}+\pi_{\nu;\mu}^{\mu}\nonumber \\
+\frac{1}{c^{2}}\left(q_{;\mu}^{\mu}u_{\nu}+q^{\mu}u_{\nu;\mu}+\theta q_{\nu}+u^{\mu}q_{\nu;\mu}\right)=0\label{eq:7}
\end{eqnarray}

\begin{equation}
nC_{n}\dot{T}+p\theta+u_{;\mu}^{\nu}\pi_{\nu}^{\mu}+q_{;\mu}^{\mu}+\frac{1}{c^{2}}\dot{u}^{\nu}q_{\nu}=0\label{eq:8}
\end{equation}
where the relation $\dot{\varepsilon}=C_{n}\dot{T}$ has been used
in order to turn the internal energy balance equation in the heat
equation. Also, we have defined $\theta=u_{;\nu}^{\nu}$ and a dot
denotes a proper time derivative. In order to complete the set of
equations, constitutive relations for the dissipative terms have to
be provided. In Ref. \cite{GRGhumo} the consitutive equations proposed
by Eckart, namely
\begin{equation}
\pi_{\nu}^{\mu}=-2\eta h_{\alpha}^{\mu}h_{\nu}^{\beta}\tau_{\beta}^{\alpha}-\zeta\theta\delta_{\nu}^{\mu}\label{eq:9}
\end{equation}
and
\begin{equation}
q^{\nu}=-\kappa h_{\mu}^{\nu}\left(T^{,\mu}+\frac{T}{c^{2}}\dot{u}^{\mu}\right)\label{eq:10}
\end{equation}
where introduced. In the present work we instead use the constitutive
relation obtained from kinetic theory by solving Botlzmann's equation
by the Chapman-Enskog approximation, that is 
\begin{equation}
q^{\nu}=-h_{\mu}^{\nu}\left(L_{TT}\frac{T^{,\mu}}{T}+L_{nT}\frac{n^{,\mu}}{n}\right)\label{eq:11}
\end{equation}
for the heat flux, while the equation for the momentum flux identical
\cite{ck}.

In order to account for self-gravitational effects prior to collapse,
we consider a Newtonian approximation of general relativity given
by the line element
\begin{equation}
ds^{2}=dr^{2}+r^{2}d\theta^{2}+r^{2}\sin^{2}\theta d\phi^{2}-\left(1-\frac{2\psi\left(r\right)}{c^{2}}\right)\left(cdt\right)^{2}\label{eq:12}
\end{equation}
where $\psi$ is the ordinary graviational potential. Since, to first
order in fluctuations, the continuity and heat equations, Eqs. (\ref{eq:6})
and (\ref{eq:8}), are only coupled with the velocity through its
divergence $\delta\theta$, it is convenient to calculate the divergence
and curl of the momentum balance equation Eq. (\ref{eq:7}). It can
be shown that by performing such calculation, the equation for the
curl of the velocity decouples from the system and has decaying solutions
for the corresponding fluctuations \cite{PhyA}. On the other hand,
the equation for $\delta\theta$ now has an inhomogeneous term that
can expressed in terms of density fluctuations by using the Poisson
equation. The resulting linearized system of equations reads

\begin{equation}
\frac{\partial}{\partial t}\left(\delta n\right)+n_{0}\delta\theta=0\label{eq:13}
\end{equation}
\begin{equation}
\tilde{\rho}_{0}\frac{\partial}{\partial t}\left(\delta\theta\right)+kT_{0}\nabla^{2}\left(\delta n\right)+n_{0}k\nabla^{2}\left(\delta T\right)-A\nabla^{2}\left(\delta\theta\right)-L_{TT}\nabla^{2}\left(\delta\dot{T}\right)-\frac{L_{nT}}{c^{2}}\nabla^{2}\left(\delta\dot{n}\right)=-4\pi Gm\tilde{\rho}_{0}\left(\delta n\right)\label{eq:14}
\end{equation}
\begin{equation}
C_{n}n_{0}\frac{\partial}{\partial t}\left(\delta T\right)+n_{0}kT_{0}\left(\delta\theta\right)-L_{TT}\nabla^{2}\left(\delta T\right)-L_{nT}\nabla^{2}\left(\delta n\right)=0\label{eq:15}
\end{equation}
where, for convenience we have defined $\tilde{\rho}_{0}=\left(n_{0}\varepsilon_{0}+p_{0}\right)/c^{2}$
and considered a fluid at rest, or an observer in a comoving frame,
such that the equilibrium value of the hydrodynamic velocity vanishes.
Equations (\ref{eq:13}-\ref{eq:15}) constitute the system of linearized
equations for the relativistic fluid to first order in the gradients
considering a weak (Newtonian) gravitational field. In the next section,
the set will be analyzed in Laplace-Fourier space in order to address
the behavior of density fluctuations.

\section{{\normalsize Dispersion Relation}}

In the previous section, the system of transport equations to first
order, both in gradients and in fluctuations has been established.
In order to analyze the behavior of density fluctuations, we proceed
by following the standard method in fluctuation theory \cite{bernepecora}.
The set is transformed to Laplace-Fourier space and the dispersion
relation is obtained by setting the corresponding determinant equal
to zero. In this case, such a relation is given by
\begin{equation}
\det\left(\begin{array}{ccc}
s & n_{0} & 0\\
-kT_{0}q^{2}+\frac{L_{nT}}{c^{2}}sq^{2}+4\pi Gm\tilde{\rho}_{0} & Aq^{2}+\tilde{\rho}_{0}s & -n_{0}kq^{2}+\frac{L_{TT}}{c^{2}}sq^{2}\\
L_{nT}q^{2} & n_{0}kT_{0} & \frac{3}{2}kn_{0}s+L_{TT}q^{2}
\end{array}\right)=0\label{eq:16}
\end{equation}
or, as a cubic equation 
\begin{equation}
s^{3}+a_{2}\left(q\right)s^{2}+a_{1}\left(q\right)s+a_{0}\left(q\right)=0\label{eq:16.5}
\end{equation}
where the coefficients are given by
\begin{equation}
a_{2}\left(q\right)=\left(\frac{A}{\tilde{\rho}_{0}}+\frac{L_{TT}}{n_{0}C_{n}}-\frac{kT_{0}L_{TT}}{\tilde{\rho}_{0}c^{2}C_{n}}-\frac{n_{0}L_{nT}}{\tilde{\rho}_{0}c^{2}}\right)q^{2}\label{eq:17}
\end{equation}
\begin{equation}
a_{1}\left(q\right)=\frac{AL_{TT}}{\tilde{\rho}_{0}c^{2}C_{n}}q^{4}+\frac{5kT_{0}n_{0}}{3\tilde{\rho}_{0}}q^{2}-4\pi Gmn_{0}\label{eq:18}
\end{equation}
and 
\begin{equation}
a_{0}\left(q\right)=\left(\frac{kT_{0}L_{TT}}{\tilde{\rho}_{0}C_{n}}-\frac{kT_{0}L_{nT}}{\tilde{\rho}_{0}C_{n}}\right)q^{4}-\frac{4\pi GmL_{TT}}{C_{n}}q^{2}\label{eq:19}
\end{equation}
It is worthwhile pointing out at this stage that by considering $A=L_{nT}=L_{TT}=0$
for the non-dissipative case, one is led directly to the Jeans wave
number 
\begin{equation}
q_{J}^{2}=\frac{4\pi Gmn_{0}}{C_{s}^{2}}\label{eq:20}
\end{equation}
where 
\begin{equation}
C_{s}^{2}=\frac{kT_{0}n_{0}}{\tilde{\rho}_{0}}\left(1+\frac{k}{C_{n}}\right)\label{eq:21}
\end{equation}
is the speed of sound in the medium, such that the only relativistic
corrections to the Jeans instability criterion neglecting dissipation
arise from the relativistic values of $\tilde{\rho}_{0}$ and $C_{n}$.
Indeed, as is verifiable from the system of equations, the relativistic
effects are mostly coupled with the dissipative ones. Since in the
non-relativistic limit $\tilde{\rho}_{0}\sim mn_{0}$ we recover the
usual Jeans wave number, where $C_{s}^{2}=\frac{5kT}{3m}$.

Turning back to the complete dispersion relation, one can extract
information about the roots of such equation by using the results
obtained in the case without gravitational field. We recall that in
such calculation, both relativistic and non-relativistic dispersion
relations have one real solution, which gives rise to a central peak
called the Rayleigh peak, and two imaginary roots which lead to damped
acustic modes corresponding to the Brillouin peaks. For a more detalied
analysis of these solutions and the corresponding scattering spectrum
the reader may consult Refs. \cite{pre09}, \cite{bernepecora} and
\cite{kyoto}. Based on those results, we here assume that the Rayleigh
peak is not significantly modified by the presence of the gravitational
field such that one of the roots of Eq. (\ref{eq:16.5}) still has
the form $s=\gamma q^{2}$ and find that this condition is satisfied
by the coefficient
\begin{equation}
\gamma=-\frac{L_{TT}}{n_{0}C_{n}}\label{eq:22}
\end{equation}
Under this assumption, we can factorize Eq. (\ref{eq:16.5}) as follows
\begin{equation}
\left(s+\gamma q^{2}\right)\left(s^{2}+\mu s+\nu\right)=0\label{eq:23}
\end{equation}
and, keeping only up to second order in $q$ we obtain
\begin{equation}
\mu=\frac{1}{\tilde{\rho}_{0}}\left(A-\frac{L_{nT}}{c^{2}}-\frac{kT_{0}L_{TT}}{c^{2}C_{n}}\right)q^{2}\label{eq:24}
\end{equation}
and
\begin{equation}
\nu=-4\pi Gmn_{0}+\frac{kT_{0}n_{0}}{\tilde{\rho}_{0}}\left(1+\frac{k}{C_{n}}\right)q^{2}\label{eq:25}
\end{equation}
The factorization in Eq. (\ref{eq:23}) corresponds to the original
dispersion relation given in Eq. (\ref{eq:16.5}) up to order $q^{2}$
which is consistent with the approximation here considered. Having
two roots of the equation, arising from a second order polynomial
yields the same physics as in the Jeans instabiliy with and without
dissipation, since one can find limiting values for $q$ for which
either oscillating or growing modes exist.

The limiting value for growing modes in density fluctations, which
physically leads to a gravitational collapse, is extracted directly
from the discriminant of the second order polynomial in Eq. (\ref{eq:16.5}),
that is
\begin{equation}
\left(A-\frac{L_{nT}}{c^{2}}-\frac{kT_{0}L_{TT}}{c^{2}C_{n}}\right)^{2}q^{4}-kT_{0}n_{0}\tilde{\rho}_{0}\left(1+\frac{k}{C_{n}}\right)q^{2}+4\pi Gmn_{0}\tilde{\rho}_{0}^{2}=0\label{eq:26}
\end{equation}
which has a root given by
\begin{equation}
q^{2}=\frac{2C_{s}^{2}\tilde{\rho}_{0}^{2}}{A\left(A-\frac{2n_{0}L_{nT}}{c^{2}}-\frac{4T_{0}L_{TT}}{3c^{2}}\right)}\left[1-\sqrt{1-\frac{4\pi G}{C_{s}^{4}\tilde{\rho}_{0}}A\left(A-\frac{2n_{0}L_{nT}}{c^{2}}-\frac{4T_{0}L_{TT}}{3c^{2}}\right)}\right]\label{eq:27}
\end{equation}
As mentioned before, this wave number corresponds to the limit between
oscillating and exponentially gowing modes. At this point it is convenient
to point out that in previous work \cite{GRGhumo}, the dispersion
relation obtained was a fourth order polynomial with a dominating
positive root which wiped out any posibility of structure formation
in the system. On the other hand, the use of Eq. (\ref{eq:11}) leads
to the same behavior as in the non-relativistic case, with small relativistic
corrections as will be shown in the next section.

\section{{\normalsize Correction to the Jeans wave number and non-relativistic
limit}}

In order to assess the magnitude of the corrections to the Jeans wave
number implied by Eq. (\ref{eq:27}) we expand the terms in the square
root, so that
\begin{equation}
q^{2}=\tilde{q}_{J}^{2}\left[1-\frac{\pi GA}{C_{s}^{4}\tilde{\rho}_{0}}\left(A-\frac{2n_{0}L_{nT}}{c^{2}}-\frac{4T_{0}L_{TT}}{3c^{2}}\right)\right]\label{eq:28}
\end{equation}
where
\begin{equation}
\tilde{q}_{J}^{2}=\frac{4\pi G\tilde{\rho}_{0}}{C_{s}^{2}}\label{eq:29}
\end{equation}
is the Jeans wave number corrected to the relativistic values of $\tilde{\rho}_{0}$
and $C_{s}^{2}$. Notice that the corrections beyond $q^{2}\sim\tilde{q}_{J}^{2}$
arise solely from the coupling of the gravitational effects with the
dissipative ones. Moreover, the relativistic corrections depend of
the heat flux but are also coupled with the viscous effect. This is
an interesting result which gives further insight in the role of viscous
dissipation in relativistic systems and reassures the need of more
research in this direction.

The non-relativistic limit is readily obtained when $c\rightarrow\infty$
since $q^{2}\rightarrow\tilde{q}_{J}^{2}$ and we obtain 
\begin{equation}
q^{2}=q_{J}^{2}\left(1-\frac{\pi GA^{2}}{C_{s}^{4}\tilde{\rho}_{0}}\right)\label{eq:30}
\end{equation}
which is precisely the result in Ref. \cite{CQG} with $q^{2}=q_{J}^{2}$
included in Eq. (\ref{eq:30}), for the case of a non-dissipative
medium. Thus, the introduction of a constitutive equation for the
heat flux in terms of gradients of scalar state variables - temperature,
density, pressure - assures that the gravitational collapse is possible
for relativistic systems following roughly the same physics as in
the non-relativistic scenario, that is with the existence of a critical
wavenumber indicating a range of values for which a structure will
form within the gas.

\section{{\normalsize Conclusions and final remarks}}

The issue of which is the correct formalism for describing dissipative
processes in relativistic hydrodynamics has recently been a subject
of intense research. It was already shown that the causality and stability
issues allegedly present in first order theories are not a concern
when the constitutive equations arising from the microscopic formalism
are considered \cite{domi2,PhyA,koide}. The causality of the system
is guaranteed as long as one considers fluctuations in the hydrodynamic
velocity even in a comoving frame \cite{domi1}. This is a completely
reasonable assumption since either a fluid at rest or an observer
in a comoving frame shall be stated as a vanishing \emph{equilibrium}
velocity. Fluctuations, which are physical evidence of the statistical
picture underlying the hydrodynamic formalism, are spontaneous and
always present. Thus, there is really no reason to consider $\delta\theta=0$.
On the other hand, the stability concern is actually more a problem
about the relaxation of such fluctuations instead of a hydrodynamic
instability \textit{per-se}. In any case, it has been shown that the
pathological behavior of the system is wiped out when gradients are
included in the momentum balance through the heat flux constitutive
equation instead of the acceleration which appears in Eckart's formalism
\cite{GRG09}. These facts lead to the firm statement that extended
theories are not \emph{strictly neccessary} and first order theories
are physically sound as long as the results of kinetic theory are
used to close the system of transport equations. However, as shown
in Ref. \cite{domi2}, introducing relaxation parameters also solves
the problem by forcing into the equations a decay that overrides the
instability. 

The question of which formalism is more adeccuate is still a subject
of debate. In the present paper the calculation presented in Ref.
\cite{GRGhumo} by two of us is carried out once again, but now using
the closure relation obtained from kinetic theory in order to argue
in favor of first order theories. It is a fact that structures form
in systems when the Jeans criterion is met. It is desirable to have
also this kind of behavior in the relativistic formalism with corrections
which vanish for mild or low temperatures. In Ref. \cite{GRGhumo}
it was clearly shown that if Eckart's closure is introduced in the
transport equations, the gravitational instability is not present
in the system, even in the non-relativstic limit. On the other hand,
what has been shown here is that by introducing the constitutive equation
obtained in Refs. \cite{degrootR,ck,PhyA}, the physics in the gravitational
collapse in a relativistic system matches the phenomenon established
by Jeans in the non-dissipative case \cite{jeans} as well as the
one described by Sandoval-Villalbazo and García-Colín for dissipative
systems \cite{CQG}. The mechanism for the onset of the gravitational
instability is the same and only slight corrections to the Jeans wave
number are obtained. It is of the authors' opinion that even if these
corrections are not astrophysically significant, the fact that the
theory predicts a collapse under particular conditions and that it
yields the correct non-relativistic limit is enough evidence to rule
it as more adequate to describe the physics of relativistic gases
than other proposals.

\end{document}